\documentclass[prd,twocolumn,floatfix,nofootinbib,showkeys]{revtex4-2}
\usepackage[utf8]{inputenc}
\usepackage[T1]{fontenc}
\usepackage{amsmath}
\usepackage{amsfonts}
\usepackage{amssymb}
\usepackage{graphicx,epstopdf}
\usepackage{datetime}
\usepackage{hyperref}
\usepackage{xcolor}
\usepackage{soul}

\graphicspath{{Figures/}}

\begin{document}	
	\title{In-medium changes of nucleon cross sections tested in neutrino-induced reactions}
	
	\author{B. Bogart}
	\affiliation{Department of Physics, University of Michigan, Ann Arbor, MI, 48109, USA}
	\email[Contact e-mail: ]{bbogart@umich.edu}
	\author{K. Gallmeister}
	\author{U. Mosel}
	\affiliation{Institut f\"ur Theoretische Physik, Universit\"at Giessen, 35392 Giessen, Germany}

	\date{\today}
	
	\begin{abstract}
	Historically studied in the context of heavy-ion collisions, the extent to which free nucleon-nucleon cross sections are modified in-medium remains undetermined by these data sets. Therefore, we investigate the impact of NN in-medium modifications on neutrino-nucleus cross section predictions using the GiBUU transport model. We find that including an in-medium lowering of the NN cross section and density dependence on $\Delta$ excitation improves agreement with MicroBooNE neutrino-argon scattering data. This is observed for both proton and neutral pion spectra in charged-current muon neutrino and neutral-current single pion production datasets. The impact of collision broadening of the $\Delta$ resonance is also investigated. The absence of $\Delta$ broadening is slightly favored, but the larger uncertainties on the pion production data prevent definitive conclusions.
	\end{abstract}
	
	
	\maketitle
	\newpage

	\section{Introduction}
	It is still an interesting problem if free nucleon-nucleon cross sections are modified when the nucleons are 
	inside a nuclear environment. For example, meson exchanges could be modified and resonance excitations could be 
	changed. Early predictions of such changes, for example those based on relativistic Brueckner-theory by Malfliet and collaborators \cite{TerHaar:1986fh}, gave a significant weakening of the NN cross sections below the resonance excitations. Li and Machleidt then took these calculations to a new level of sophistication by using the Bonn meson-exchange model \cite{Li:1993rwa,Li:1993ef}. 
	Since then many investigations have been performed, partly for testing the Li-Machleidt cross sections and partly for extracting the nuclear medium effects from comparisons to experimental data. No clear picture has emerged from these studies, partly because in heavy-ion collisions, mostly used for these investigations of in-medium effects, one deals with a non-equilibrium situation and signals that are integrated over the time-development of the collision (see \cite{Henri:2020ezr} and Refs. therein). In addition they are sensitive to the momentum-dependence of the NN interactions at momenta above the Fermi-momentum. There, however, nucleon resonance excitations with their own changes set in. Other studies have used the sensitivity of the width of the giant resonance to in-medium changes of the NN cross section \cite{DiToro:1998ki,Wang:2020xgk}.
	
	There is a class of reactions that has not been used to explore the NN in-medium changes. These are lepton induced reactions such as ($e,e'p$) or ($\nu, \mu p$). In these reactions the nuclear target is at rest and the densities are restricted to be between 0 and the nuclear saturation density $\rho_0$ with most of the target nucleons sitting at about $2/3 \rho_0$. Measurements of electron-induced reactions have often aimed at getting information on the bound nucleons and have tried to minimize the final state interactions of the knocked-out proton~\cite{PhysRevD.105.112002,PhysRevD.103.034604,Nature.560.617621,PhysLettB.608.4752}. This reduces their sensitivity to in-medium modifications of NN cross sections. We have, therefore, looked at the impact of NN in-medium changes in neutrino-induced reactions using recent neutrino-argon scattering data from MicroBooNE which provide ejected proton multiplicities, proton spectra, and neutral pion spectra. The purpose of this letter is to explore the sensitivity of these observables to possible in-medium changes.

 The authors of Ref.\ \cite{Lu:2019nmf} looked at nuclear effects in neutrino interactions on nuclei in the MINERvA and T2K experiments by analyzing proton transverse kinetic imbalances, but did not address the question of possible in-medium NN effects on the observables nor compare with data from these experiments.
 In the MINERvA experiment the energy is fairly high. This means many different processes can take place in the final state interaction phase and resonance excitations, with their own in-medium changes, prevail. This is not so in the MicroBooNE experiment~\cite{uboone_detector} which we investigate in the present paper. In this experiment, because of the lower beam energies~\cite{Phys.Rev.D.79.072002}, only quasielastic scattering, meson exchange processes and $\Delta$ excitation are relevant in the initial interaction (ISI) and the open channels during the final state interactions (FSI) are limited. Furthermore, the MicroBooNE liquid argon time projection chamber detector is able to track protons down to low kinetic energies with good precision, with lower detection thresholds than T2K and MINERvA, which enables spectra of these particles to be measured over a wide range of kinetic energies spanning from 10s up to 1000 MeV. The situation is similar for neutral pions, which are detected via the identification of their two decay photons, and thus have no minimum kinetic energy for acceptance. These low thresholds make MicroBooNE data an ideal place to investigate the impact of NN in-medium changes, which will have a more prominent effect on the production of low energy protons and pions. 
    
\section{Method}
This investigation is using the GiBUU theory framework and code in its 2023 version; details of the underlying theory and practical implementation are given in Refs. \cite{Buss:2011mx,Mosel:2023zek}. The code is available for download from \textit{gibuu.hepforge.org}.

The in-medium NN interactions for the FSI which are implemented in GiBUU are those of Li and Machleidt \cite{Li:1993rwa,Li:1993ef}. They lower the elastic NN cross sections as a function of density. In addition the inelastic resonance excitation in NN collisions is modified by using the in-medium change explored by of Song and Ko \cite{Song:2015hua}
\begin{equation}
	\sigma_{NN \to N \Delta} (\rho) = \sigma_{NN \to N \Delta} (0)\, {\rm exp}(-1.2\, \rho/\rho_0)~;
\end{equation}
 it decreases the excitation of the $\Delta$ resonance with increasing density. GiBUU treats this consistently in both the pion production and the pion absorption channels. In the present study it mainly affects the pion absorption through the FSI process $\pi N \to \Delta, \Delta N \to NN$. 

Another in-medium change is due to the collisional broadening of the $\Delta$ resonance. This plays a role mainly in the initial neutrino-nucleon interactions because secondary $\Delta$ excitations during the FSI are very rare at the low energy of the MicroBooNE. For this collisional broadening GiBUU has implemented the density- and momentum-dependent collisional width for the $\Delta$ obtained by Oset and Salcedo \cite{Oset:1987re}. We note that the parametrization given there contains some uncertainty as discussed, for example, by the authors of Ref. \cite{Gonzalez-Jimenez:2017fea}.

We take all the in-medium changes directly from the papers quoted; there is no tuning involved. We also mention here that GiBUU does not contain any coherent processes. This has a negligible impact on the muon-neutrino charged current ($\nu_\mu \text{CC}$) distributions shown in Sec. III A, but could give a modest yet non-negligible increase to the predicted $\pi^0$ spectra shown in Sec. III B.
   
\section{Comparison with experiment}
Ref.\ \cite{MicroBooNE:2024xod} contains an extensive set of $\nu_\mu \text{CC}$ cross section measurements on argon from the MicroBooNE experiment. These measurements are inclusive, but divide the channel into final states with and without protons above the detector's 35 MeV kinetic energy tracking threshold. These are referred to as the ``Np'' and ``0p'' final states, respectively. Cross sections as a function of the proton kinematics and proton multiplicity of the final state are also presented. The work includes a number of comparisons of the MicroBooNE data to the ``standard'' GiBUU version, i.e. the one without any in-medium corrections. Similar to other MicroBooNE measurements~\cite{Phys.Rev.Lett.131.101802.2023,Phys.Rev.Lett.128.151801.2022,arXiv2403.19574}, GiBUU demonstrates a consistent ability to describe the data. 

Here we now enhance this comparison by showing also the results obtained with the in-medium corrections. Because the results in Ref.\ \cite{MicroBooNE:2024xod} include the 0p topology, which is rich in quasielastic interactions where the proton produced in the ISI was significantly impacted by FSI, this data set is expected to be particularly sensitive to in-medium modifications. As suggested in Ref.\ \cite{WSVD}, we account for any bias induced by the regularization of the data in cross section extraction by multiplying our predictions by the $A_C$ matrix reported by the experiment. We also quantify our predictions' agreement with the data via $\chi^2$ values computed using the reported covariance matrix, which includes correlations between all measurement bins, including pairs of bins belonging to different distributions.

\subsection{Proton spectra}
	\begin{figure}
		\centering
		\includegraphics[width=0.9\linewidth]{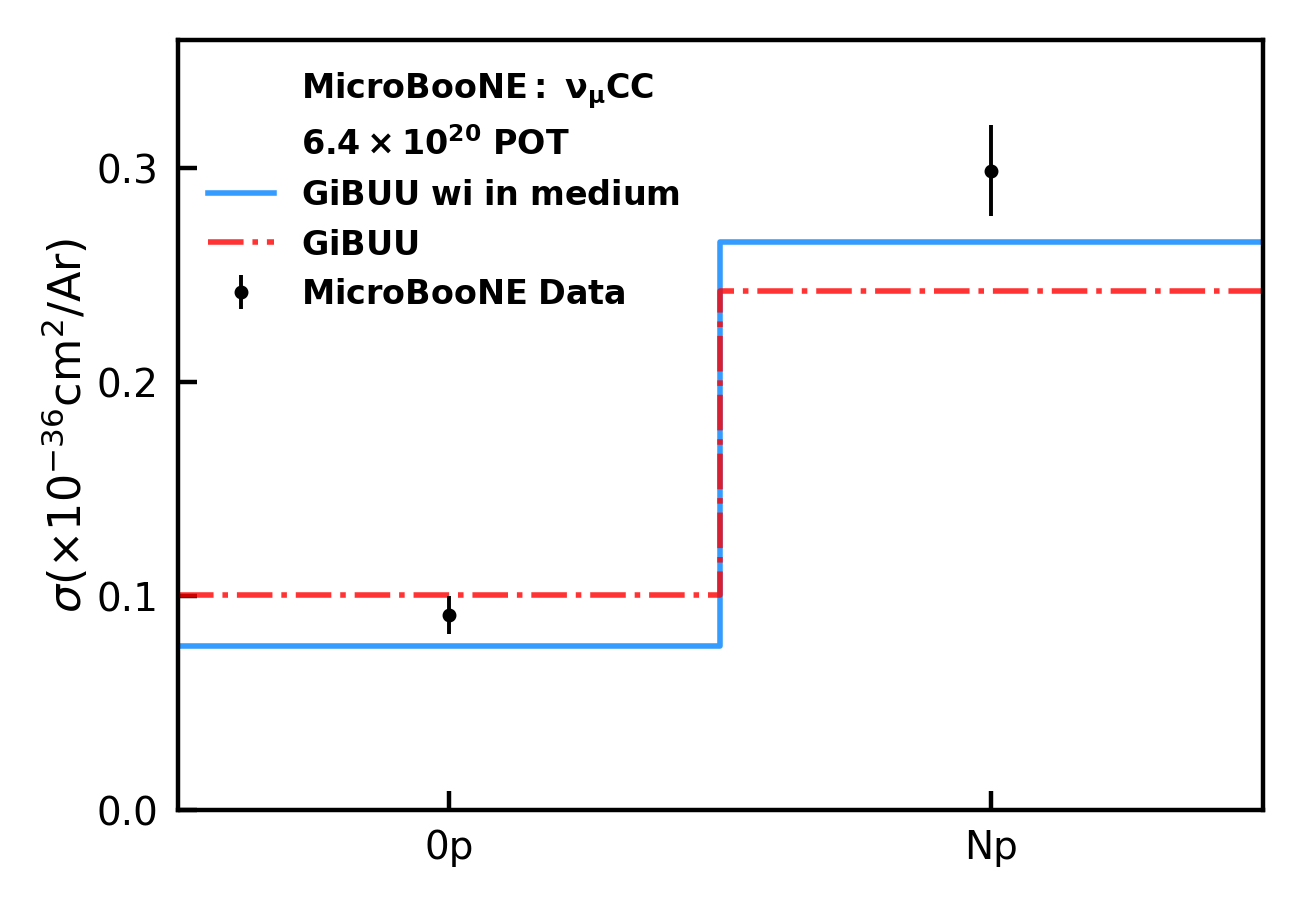}
		\caption{Total 0p and Np $\nu_\mu \text{CC}$ cross sections measured by MicroBooNE in Ref.\ \cite{MicroBooNE:2024xod} compared to GiBUU predictions with and without in-medium modifications to the NN cross sections.}
		\label{fig:total}
	\end{figure}	
	
 In Fig.\ \ref{fig:total} we first show the total $\nu_\mu \text{CC}$ cross sections for 0 proton and N proton (where N\ $\geq1$) final states. In order to be counted the protons have to have a kinetic energy of >~35~MeV. We include a prediction with (``GiBUU wi in medium'') and without (``GiBUU'') the Li-Machleid and Song-Ko and NN modifications. The Oset-Salcedo $\Delta$ broadening is excluded for both predictions. Overall, the effect of the in-medium corrections is not dramatic. The 0p cross section is somewhat lowered when switching on the in-medium corrections and it is somewhat enhanced for the Np events. This behavior is a direct consequence of the in-medium lowering of the NN cross sections. The naive final state for a quasielastic $\nu_\mu \text{CC}$ interaction contains one muon and one proton. Thus, because quasielastic interactions dominate at MicroBooNE energies, the magnitude of the 0p channel is primarily driven by final state interactions in which the proton looses some of its kinetic energy and ends up either bound in the target nucleus or below the experimental tracking threshold. Protons can also get lost all together by a charge-changing pn collision. When the NN cross sections are lowered in medium there are less of these collisions which leads to less shift in strength from Np to 0p, as observed in Fig.\ \ref{fig:total}. 
 
	\begin{figure}
		\centering
		\includegraphics[width=0.9\linewidth]{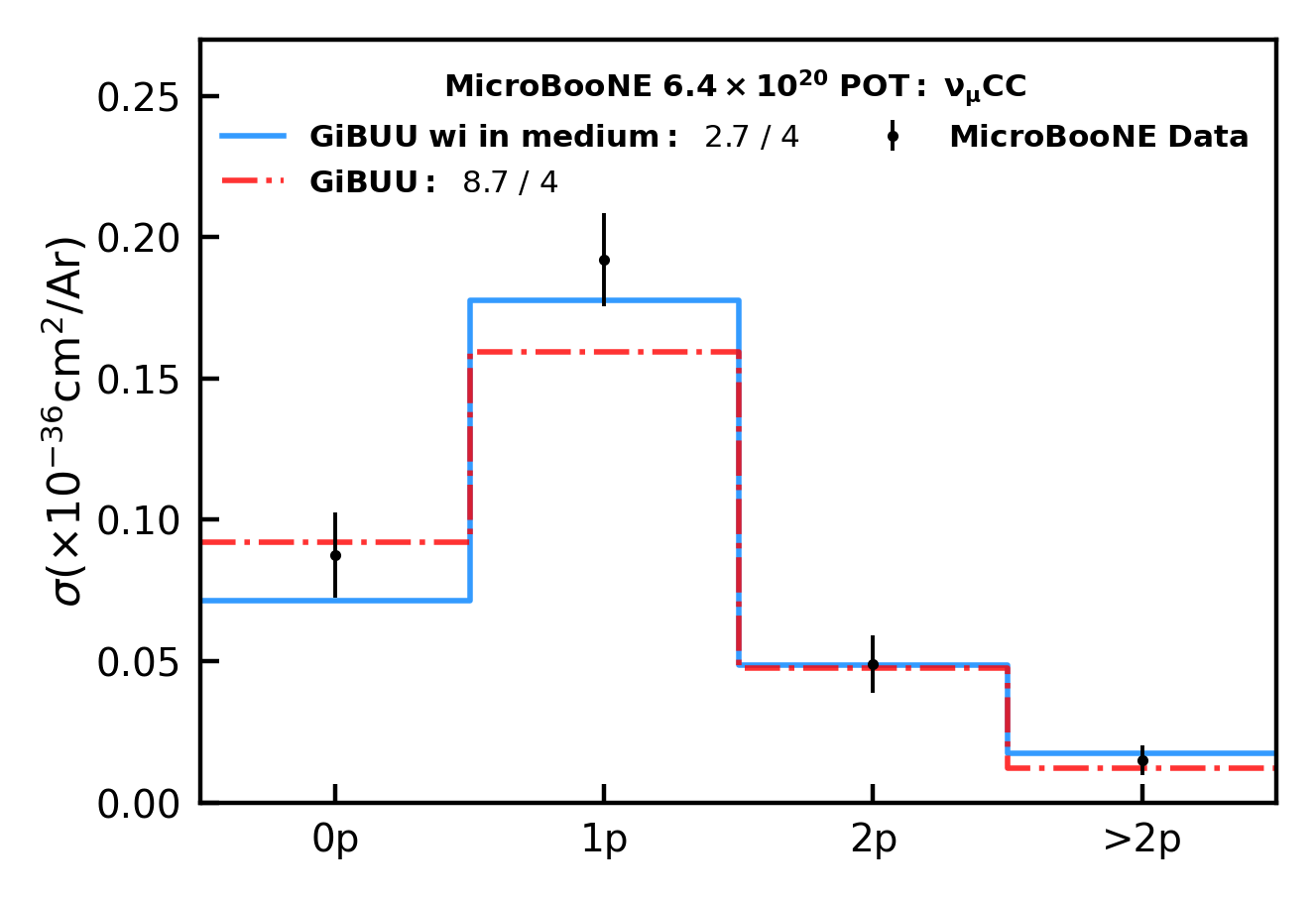}
		\caption{Proton multiplicity in $\nu_\mu \text{CC}$ interactions measured by MicroBooNE in Ref.\ \cite{MicroBooNE:2024xod} compared to GiBUU predictions with and without in-medium modifications to the NN cross sections. The $\chi^2/$bin in the legend is calculated with the covariance matrix reported by the experiment.}
		\label{fig:multipl}
	\end{figure}
 
 This is also reflected in the multiplicity distribution shown in Fig.\ \ref{fig:multipl}. As in Fig.\ \ref{fig:total}, protons only count towards the multiplicity if they have a kinetic energy of >~35~MeV and the small differences between the values in the 0p bins are attributable to slightly different experimental cuts and efficiency corrections. Here strength is shifted from 0p to mostly 1p, with only small effects in the 2p and $>$2p bins. This is unsurprising given the dominance of quasielastic scattering at MicroBooNE energies. The shift in strength is favored by the data and allows the in-medium prediction to fall within 1$\sigma$ of the measurement on all multiplicity bins. The computed $\chi^2/$bin value reflects this improved agreement and decreases from 8.7/4 to 2.7/4 when the in-medium corrections are turned on.
 
 	\begin{figure}
 	\centering
 	\includegraphics[width=0.9\linewidth]{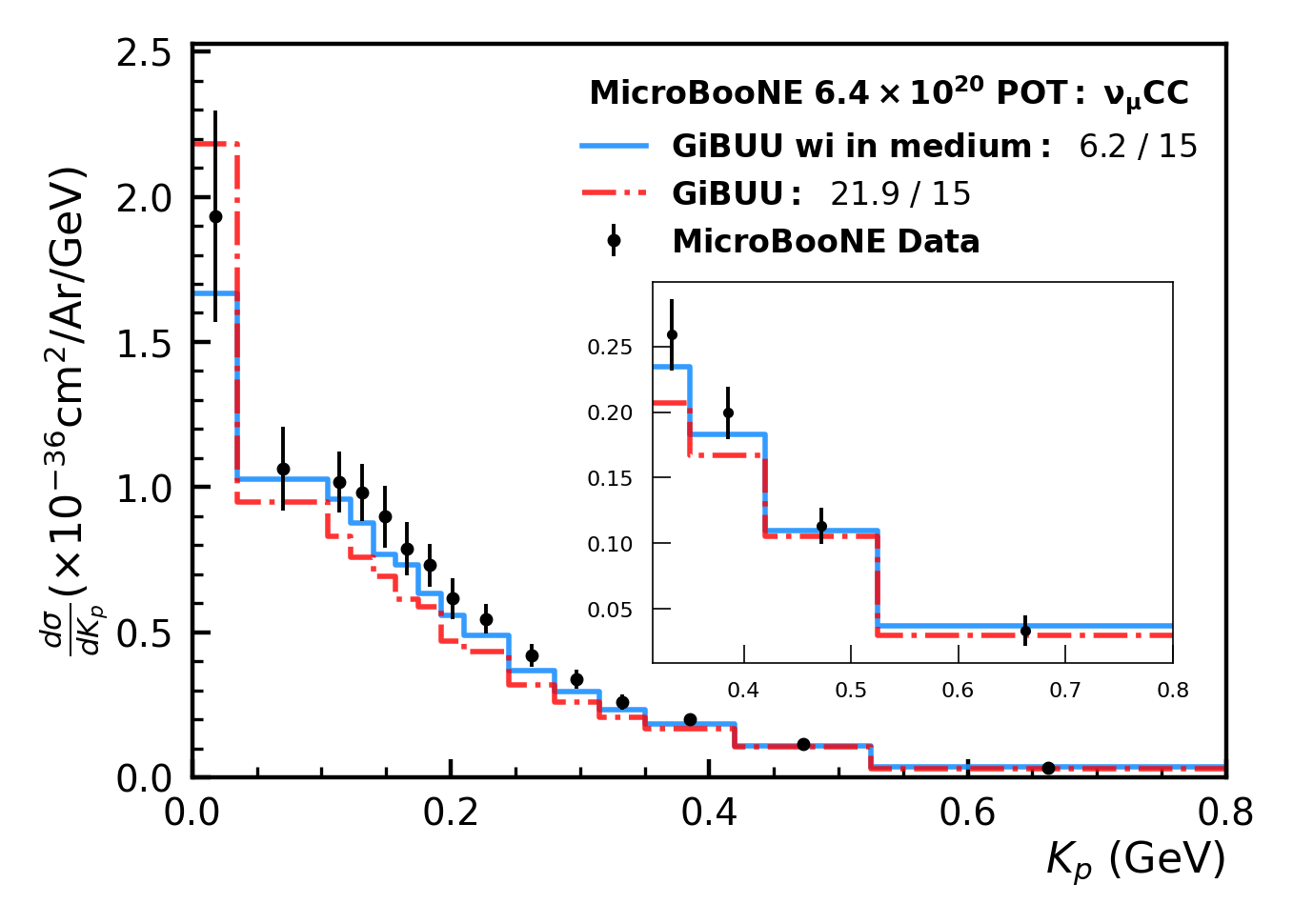}
 	\caption{Same as Fig.\ \ref{fig:multipl}, but for the differential cross section as a function of the most energetic proton's kinetic energy. The first bin includes interactions in which no proton leaves the nucleus. The inset shows a magnified view of the high energy bins.}
 	\label{fig:Tp}
 \end{figure}
	
In Fig.\ \ref{fig:Tp} we show the kinetic energy distribution for the most energetic final state proton in $\nu_\mu \text{CC}$ reactions. The first bin of this distribution includes all interactions without a final state proton and interactions with a final state proton below the 35~MeV tracking threshold. The cross section is significantly decreased for protons with energies below the threshold when the in-medium effects are turned on and the overall distribution is markedly enhanced in much better agreement with experiment; the $\chi^2$ decreases from 21.9/15 down to 6.2/15.

	    \begin{figure}
			\centering
			\includegraphics[width=0.9\linewidth]{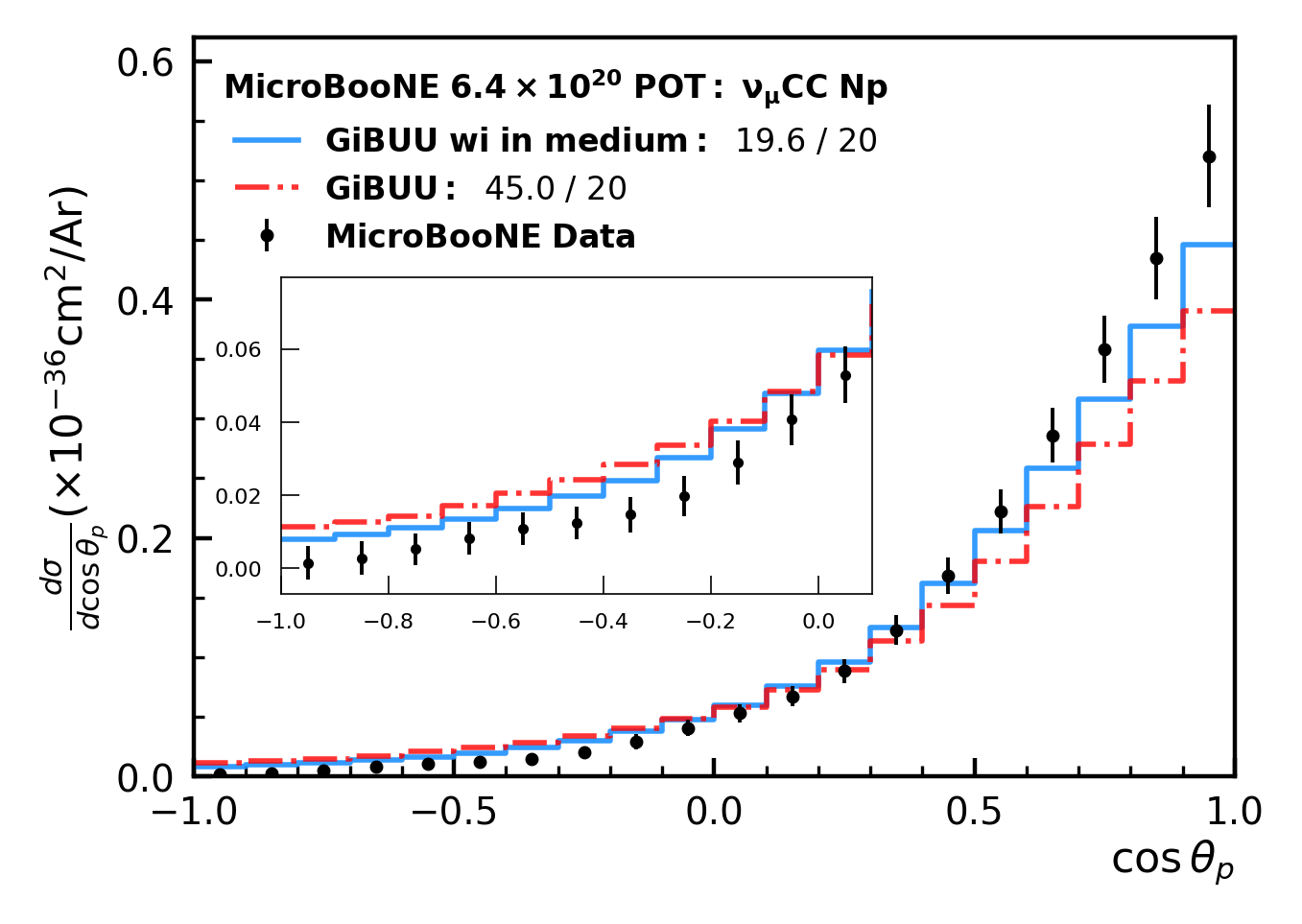}
			\caption{Same as Fig.\ \ref{fig:multipl}, but for the differential cross section as a function of the most energetic proton's scattering angle. The inset shows a magnified view of the backwards bins.}
			\label{fig:costheta}
		\end{figure}
  
Finally, in Fig.\ \ref{fig:costheta} we show the angular distribution of the most energetic outgoing proton. The scattering angle is defined with respect to the incoming neutrino beam. The calculation using the free NN cross sections underestimates the distribution at forward angles and overestimates it at backwards angles. When the in-medium changes are turned on the forward behavior is improved. This is a direct consequence of the smaller NN cross sections which leads to less re-scattering to perpendicular and backwards angles. 

\subsection{$\pi^0$ spectra}
We now also look for effects of in-medium changes of the NN cross sections in neutral current production of uncharged pions. For this investigation, we compare our predictions with MicroBooNE neutral current single pion production (NC$\pi^0$) data from Ref.\ \cite{arXiv2404.10948}. At MicroBooNE energies most of the pions are produced via the $\Delta$ resonance so these cross sections are sensitive both to the properties of the $\Delta$ resonance \cite{Oset:1987re} in the ISI and to the resonance suppression proposed in Ref.\ \cite{Song:2015hua} in the FSI.  As in Ref.\ \cite{MicroBooNE:2024xod}, Ref.\ \cite{arXiv2404.10948} reports $A_C$ matrices, which we use to account for the regularization of the data, as well as a covariance matrix that includes correlations between all measurement bins, which we use to quantify agreement with the data. For these comparisons, we note that GiBUU does not simulate coherent pion production, for which there is data that indicates the process could make up around 10-20\% of the total cross section at these energies~\cite{Phys.Rev.D.81.111102,Phys.Lett.B.664.41}.

\begin{figure}[h]
	\centering
	\includegraphics[width=0.9\linewidth]{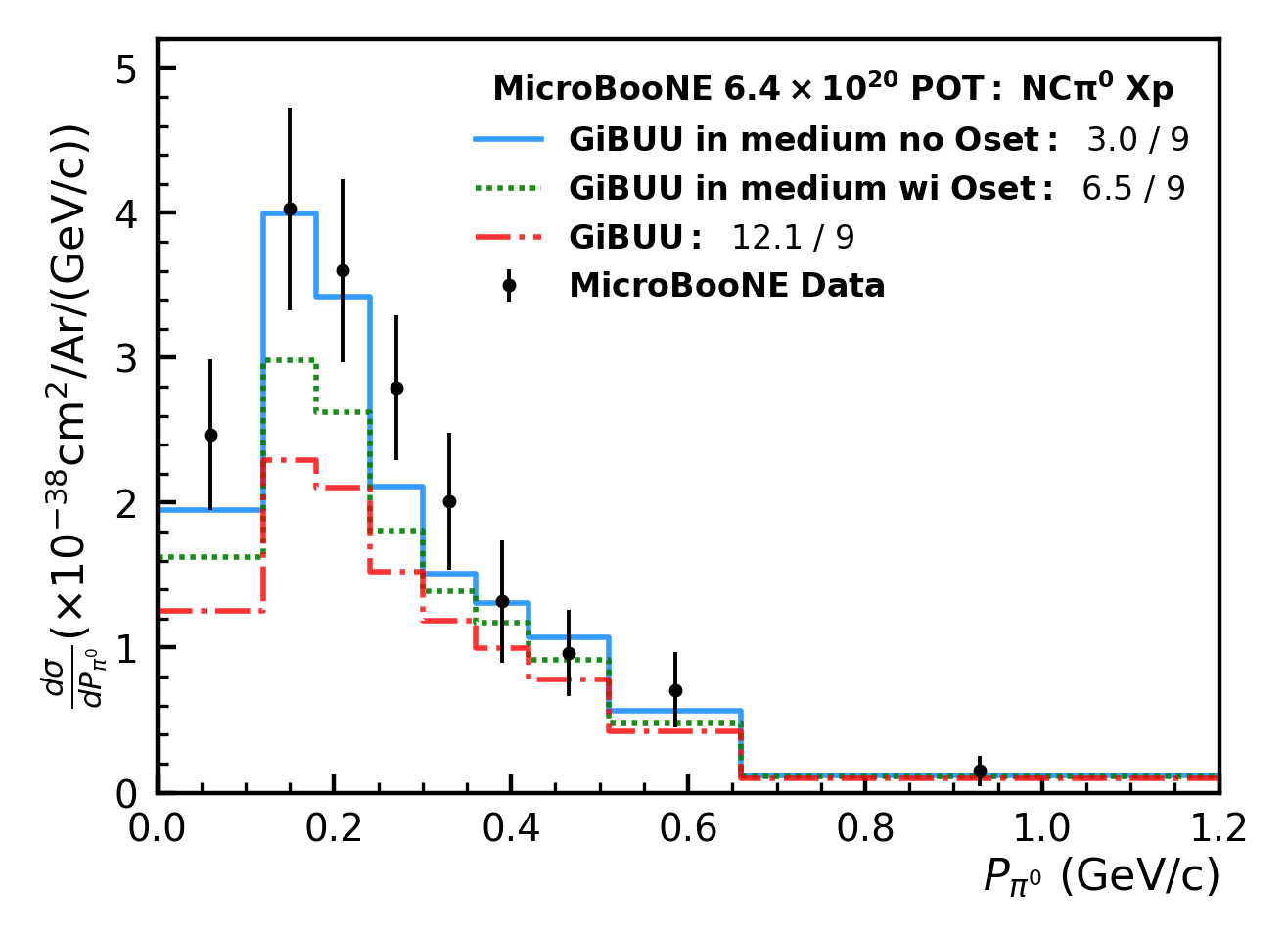}
	\caption{The $\pi^0$ momentum spectrum for neutral current interactions that produce any number of protons and a single neutral pion. Data from Ref.~\cite{arXiv2404.10948} are compared to predictions that use free NN cross sections (``GiBUU''), use in-medium NN cross sections and the free $\Delta$ width (``GiBUU in medium no Oset''), and use in-medium NN cross sections and the Oset broadening of the $\Delta$ resonance (``GiBUU in medium wi Oset''). The $\chi^2/$bin in the legend is calculated with the covariance matrix reported by the experiment.}
	\label{fig:ppi0xp}
\end{figure}

The momentum spectrum of uncharged pions produced in neutral-current interactions with any number of final-state protons is shown in Fig.\ \ref{fig:ppi0xp}. Here the influence of the in-medium corrections is significant, in particular at the peak of the distribution. This figure shows two curves obtained with in-medium changes. In the curve ``GiBUU~in~medium~wi~Oset'' the Li-Machleidt and Song-Ko in-medium changes are turned on and the broadening of the $\Delta$ resonance given by the parametrization of Oset et al. \cite{Oset:1987re} is included. In the curve ``GiBUU~in~medium~no~Oset'', we include the Li-Machleidt and Song-Ko in-medium changes, but turn off $\Delta$ broadening. 

Including the Song-Ko density dependence for $\Delta$ excitation increases the NC$\pi^0$ cross section by reducing $\Delta$ absorption primarily through $\Delta N \rightarrow NN$, which is also impacted due to detailed balance. This decreases the probability that the $\Delta$ is absorbed into the nucleus before it can decay to a pion, thereby increasing the $\pi^0$ yield. This is seen clearly in Fig.~\ref{fig:ppi0xp}. The increase in strength around the peak of the distribution is favored by the data and results in improved shape and normalization agreement for the ``GiBUU in medium no Oset'' prediction, which includes these in-medium corrections. Including the Oset collision broadening leads to a lower peak value of the momentum distribution, with very little effect on the tail. This is a consequence of the increase in the effective $\Delta$ width, which lowers the initial pion production \cite{Lalakulich:2012cj}. Though not appearing to provide better agreement with the data, the ``GiBUU in medium wi Oset'' prediction's underestimation of the data at lower momenta may be attributable to coherent pion production, which is not accounted for in these predictions. While there is some evidence that $\Delta$ broadening is necessary to describe photo- and electro-nuclear reactions~\cite{Ericson:1988gk} the validity of the special Oset-Salcedo in-medium width remains to be seen. Indeed, both the predictions with and without this broadening provide a reasonable description of the data, especially considering that the contribution to the cross section from neutral current coherent pion production is unknown and yet to be measured on argon.

\begin{figure}
	\centering
	\includegraphics[width=0.9\linewidth]{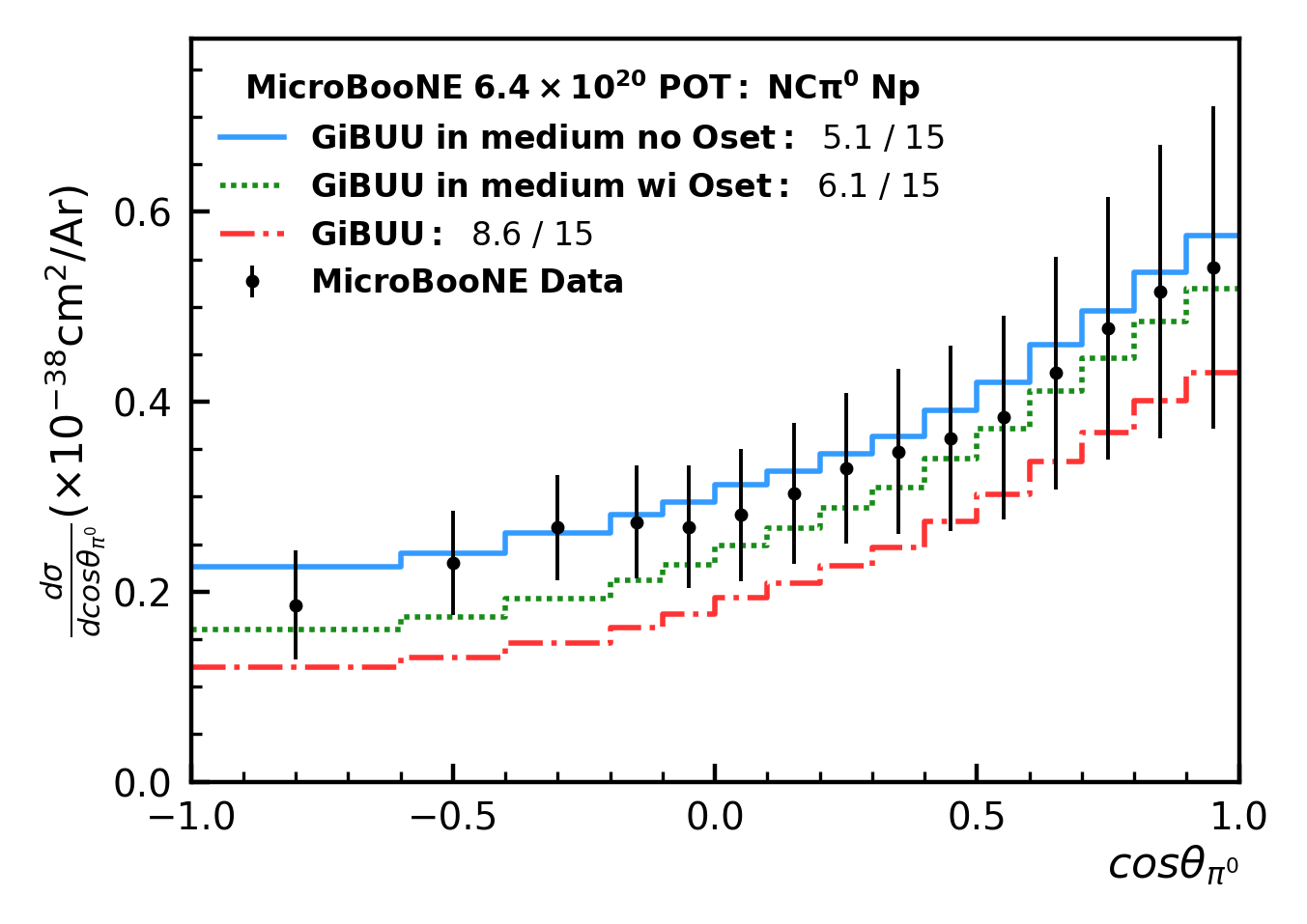}
	\caption{The $\pi^0$ angular spectrum for neutral current interactions that produce at least one proton and a single neutral pion. Data are from \cite{arXiv2404.10948}. See Fig.\ \ref{fig:ppi0xp} for additional details.}
	\label{fig:cospi0np}
\end{figure}

In Fig.\ \ref{fig:cospi0np} we show the scattering angle of the $\pi^0$ for neutral current interactions that also produce a proton above the 35~MeV kinetic energy tracking threshold of the MicroBooNE detector. The in-medium corrections slightly improve the agreement with experiment and the overall angular distribution is described quite well. Here, the data is not sensitive to the in-medium change of the $\Delta$ spectral function given in \cite{Oset:1987re}. The predictions with and without the Oset broadening both fall well within the uncertainties of the measurement. 

\begin{figure}
	\centering
	\includegraphics[width=0.9\linewidth]{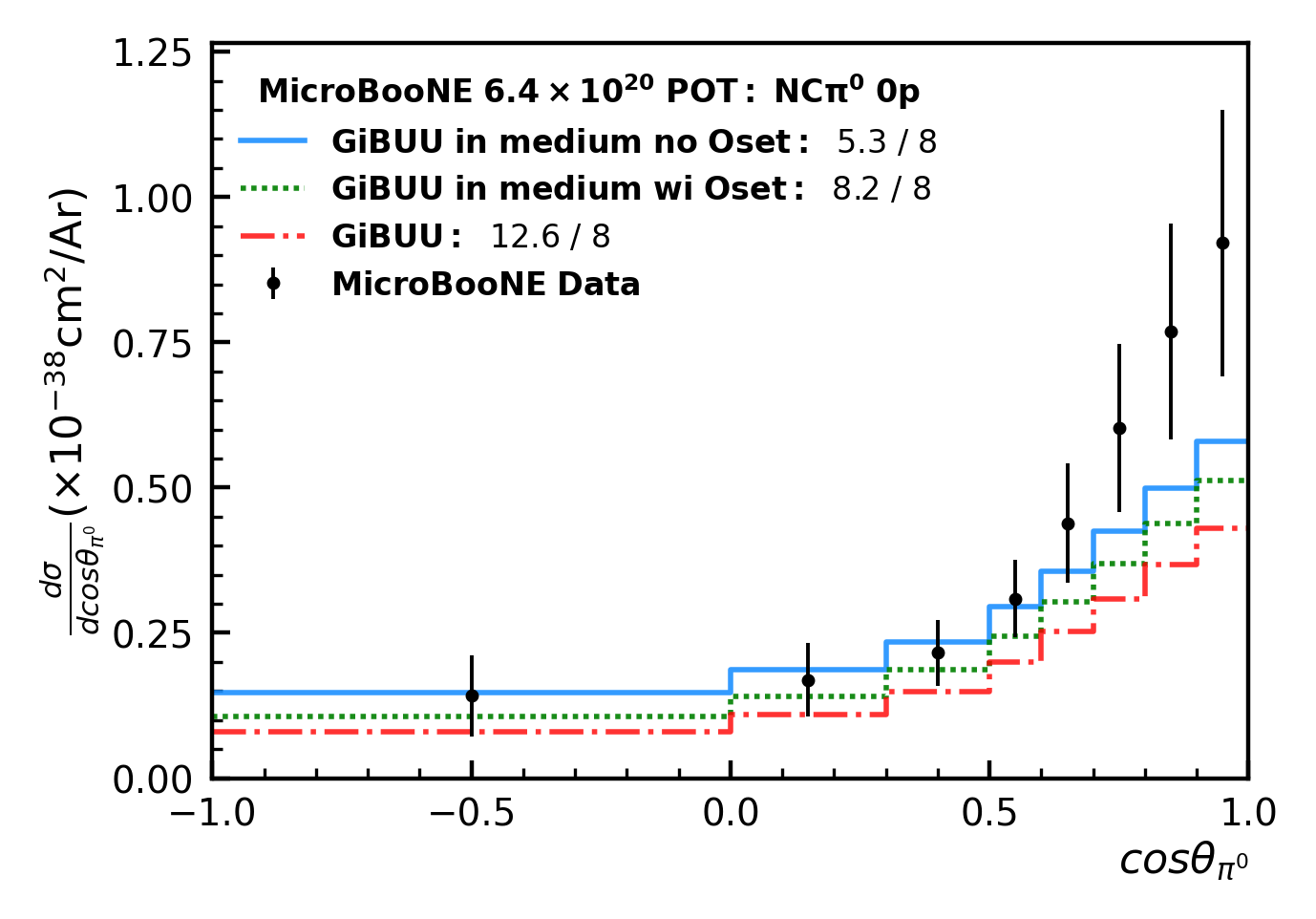}
	\caption{The $\pi^0$ angular spectrum for neutral current interactions that produce no final state protons and a single neutral pion. Data are from \cite{arXiv2404.10948}. See Fig.\ \ref{fig:ppi0xp} for additional details.}
	\label{fig:cospi00p}
\end{figure}

Finally, in Fig.\ \ref{fig:cospi00p} we give the same angular distribution for $\pi^0$s, but now for events with 0 protons in the acceptance range. While turning on the in-medium changes again improves the agreement with experiment, there is still a clear underestimate of the cross section at forward angles ($\cos\theta_{\pi^0} > 0.7$). We attribute this residual $\pi^0$ yield above the predictions as being at least partially due to coherent pion production, which is not contained in GiBUU. Integrating the difference between the GiBUU curves and the data gives a value for the coherent cross section that is in the correct range with respect to predictions from the NEUT 5.4.0.1~\cite{neut} and GENIE v3.0.6 G18\_10a\_02\_11a~\cite{genie} neutrino event generators, which estimate that the cross section for the process on argon is $8.6\times10^{-40}$~cm$^2$/Ar and $3.7\times10^{-40}$~cm$^2$/Ar, respectively, and measurements of coherent pion production at similar energies~\cite{Phys.Rev.D.81.111102,Phys.Lett.B.664.41}. These measurements are, however, on different nuclear targets so we caution against a direct comparison due to possible differences in the A scaling for resonant and coherent processes. Alternatively, this residual 0p deficit may also suggest a need for further in-medium modifications to pn cross sections which may have a different influence on the 0p to Np ratio in NC$\pi^0$ production than in the $\nu_\mu$CC channel due to the initial neutral current interaction being capable of producing a neutron. 

\subsection{Global Comparison}
Beyond the results shown here, we also compared the GiBUU predictions with and without the in-medium modifications to the rest of the $\nu_\mu$CC measurements reported in Ref.\ \cite{MicroBooNE:2024xod}. The in-medium predictions show comparable or improved agreement with the data in all cases. To quantify this, for both predictions, we compute a $\chi^2$ value across the entirety of measurements reported in Ref.\ \cite{MicroBooNE:2024xod}. Because ``blockwise unfolding''\ \cite{blockwise} was employed in Ref.\ \cite{MicroBooNE:2024xod}, these $\chi^2$ values account for the correlations between all measurement bins, including those corresponding to different variables. The GiBUU prediction with the free NN cross sections achieves a $\chi^2/$bin of 1064/704. Including the in-medium modification lowers this to 795/704. Similarly, for the three measurements we show in Figs.\ \ref{fig:multipl}-\ \ref{fig:costheta}, the $\chi^2/$bin decreases from 69.8/39 to 24.7/39, indicating a significant improvement in the description of this data when in-medium modifications of Li-Machleidt and Song-Ko are included.

We performed the analogous comparison for the rest of the NC$\pi^0$ results presented in Ref.\ \cite{arXiv2404.10948}, which also reports correlations between all measurement bins. Here, including the Li-Machleidt and Song-Ko in-medium changes improves agreement for all results. In particular, comparing the ``GiBUU'' prediction to the ``GiBUU~in~medium~no~Oset'' prediction, the $\chi^2/$bin for the double-differential measurement of the $\pi^0$ angle and momentum improves from 20.3/24 to 9.3/24. The full blockwise results are less sensitive to these differences and only reduces from 43.0/78 to 31.9/78. For this comparison, we have excluded the 0p bins at low momentum and forward angles. The residual deficit in these bins, which we attribute partly to coherent pion production, becomes the primary driver of the $\chi^2$ value. This renders the test statistic insensitive to changes throughout the rest of phase space, hence their exclusion. 

These comparisons paint a consistent picture across both sets of measurements; the in-medium modification of Li-Machleidt and Song-Ko improves agreement with experiment. Though the initial interaction in the NC$\pi^0$ channel differs from the $\nu_\mu$CC channel in that it is dominated by resonant interactions rather than quasielastic interactions and mediated by the neutral current rather than the charged current, the FSI experience by the outgoing hadronic reaction products are identical. As such, it is reassuring to see that the treatment of in-medium effects improves the description of data in both channels. Moreover, the Supplemental Material of another MicroBooNE result~\cite{arXiv:2404.09949}, which measured $\pi^0$ production in charged-current interactions, has also included comparisons to GiBUU with these in-medium modifications. Those comparisons are consistent with what we show here. The in-medium prediction better describes that data, particularly for the $\pi^0$ momentum, which shows analogous trends to Fig.\ \ref{fig:ppi0xp}.

\section{Conclusion}			
The degree to which free nucleon-nucleon cross sections are modified within the nucleus remains an open question. Theoretical investigations suggest that there is a significant reduction of NN cross section in medium, but experimental investigations, mostly from heavy-ion collisions, has been unable to provide further insight. However, lepton induced reaction on at-rest nuclear targets are yet to be thoroughly explored. In particular, the growing body of cross section measurements being made on heavy nuclear targets by accelerator-based neutrino experiments has not been examined in this light. The well-established need for a robust description of the final state interactions that the initial reactant products experience as they exit the nucleus suggests that these measurements will be sensitive to modifications of nucleon-nucleon forces within the nucleus.

As such, we have utilized the GiBUU model to probe MicroBooNE neutrino-argon scattering data for sensitivity to in-medium modifications of NN cross sections. Using GiBUU, we implemented the lowering of the NN cross sections according to the work of Li and Machleidt. We also account for the density dependence of the $\Delta$ excitation cross section according to the work of Song and Ko. Comparing these predictions and ones without in-medium modifications against MicroBooNE $\nu_\mu$CC and NC$\pi^0$ measurements reveals that these data are quite sensitive in-medium effects. Including the in-medium lowering of the NN cross sections and density dependence of $\Delta$ excitation better reproduces the measured proton and $\pi^0$ spectra. This is especially apparent at low proton and $\pi^0$ energies, which are regions of phase space  significantly impacted by FSI. Overall, our investigation indicated that accounting for these modification within the nuclear medium is essential in obtaining a satisfactory description of the data. 

For the NC$\pi^0$ data we also examine the impact of collision broadening of the $\Delta$ resonance. Excluding the Oset-Salcedo density- and momentum-dependent broadening for the $\Delta$ is slightly favored, but some uncertainties in the collisional broadening used and the larger uncertainties on the pion production data prevents definitive conclusions. The lack of coherent pion production in GiBUU and relative sparsity of experimental data for this channel further complicates the interpretation of these results and a dedicated measurement of the coherent process on argon by MicroBooNE or other
experiments in the Short-Baseline Neutrino (SBN) program could shed additional light on this disagreement at forward angles. Moreover, the neutrino data we investigate are not directly sensitive to $\Delta$ broadening, only to a lowering of the maximum of the $\Delta$ mass distribution and, consequently, its cross section, which is also sensitive to the $\Delta$ potential inside the nucleus. The clearest way to disentangle these effects would be through analyzing the invariant mass distributions of $\pi$N pairs. Such a measurement could also be performed by MicroBooNE or other SBN experiments. 

A measurement of neutron yields in neutrino-argon interactions using liquid argon time projection chamber neutron detection techniques~\cite{arXiv:2406.10583} would also be interesting as it may allow one to explore the need for in-medium modification from a different angle and could shed additional light on a possible need for further modifications to pn cross sections. Beyond neutrino experiments, higher energy data on di-hadron production on nuclei obtained with a positron beam \cite{HERMES:2005mar} have shown that such data are sensitive to the hadron formation process and to the final-state interactions. At lower energies, such experiments with electron or neutrino beams could thus provide additional information on the nature of NN interactions experienced by the final state particles.

\begin{acknowledgments}
BB is supported by the Department of Energy, Office of Science, under Award No. DE-SC0007859.
\end{acknowledgments}


\begin{thebibliography}{0}%
\makeatletter
\providecommand \@ifxundefined [1]{%
 \@ifx{#1\undefined}
}%
\providecommand \@ifnum [1]{%
 \ifnum #1\expandafter \@firstoftwo
 \else \expandafter \@secondoftwo
 \fi
}%
\providecommand \@ifx [1]{%
 \ifx #1\expandafter \@firstoftwo
 \else \expandafter \@secondoftwo
 \fi
}%
\providecommand \natexlab [1]{#1}%
\providecommand \enquote  [1]{``#1''}%
\providecommand \bibnamefont  [1]{#1}%
\providecommand \bibfnamefont [1]{#1}%
\providecommand \citenamefont [1]{#1}%
\providecommand \href@noop [0]{\@secondoftwo}%
\providecommand \href [0]{\begingroup \@sanitize@url \@href}%
\providecommand \@href[1]{\@@startlink{#1}\@@href}%
\providecommand \@@href[1]{\endgroup#1\@@endlink}%
\providecommand \@sanitize@url [0]{\catcode `\\12\catcode `\$12\catcode `\&12\catcode `\#12\catcode `\^12\catcode `\_12\catcode `\%12\relax}%
\providecommand \@@startlink[1]{}%
\providecommand \@@endlink[0]{}%
\providecommand \url  [0]{\begingroup\@sanitize@url \@url }%
\providecommand \@url [1]{\endgroup\@href {#1}{\urlprefix }}%
\providecommand \urlprefix  [0]{URL }%
\providecommand \Eprint [0]{\href }%
\providecommand \doibase [0]{https://doi.org/}%
\providecommand \selectlanguage [0]{\@gobble}%
\providecommand \bibinfo  [0]{\@secondoftwo}%
\providecommand \bibfield  [0]{\@secondoftwo}%
\providecommand \translation [1]{[#1]}%
\providecommand \BibitemOpen [0]{}%
\providecommand \bibitemStop [0]{}%
\providecommand \bibitemNoStop [0]{.\EOS\space}%
\providecommand \EOS [0]{\spacefactor3000\relax}%
\providecommand \BibitemShut  [1]{\csname bibitem#1\endcsname}%
\let\auto@bib@innerbib\@empty
\end{thebibliography}%


\begin{thebibliography}{99}
	
	\bibitem{TerHaar:1986fh}
	B.~Ter Haar and R.~Malfliet,
	Phys. Lett. B \textbf{172}, 10-16 (1986)
	doi:10.1016/0370-2693(86)90207-8.
	
	
	\bibitem{Li:1993rwa}
	G.~Q.~Li and R.~Machleidt,
	Phys. Rev. C \textbf{48}, 1702 (1993)
	doi:10.1103/PhysRevC.48.1702
	[arXiv:nucl-th/9307028 [nucl-th]].	
	
	\bibitem{Li:1993ef}
	G.~Q.~Li and R.~Machleidt,
	Phys. Rev. C \textbf{49}, 566 (1994)
	doi:10.1103/PhysRevC.49.566
	[arXiv:nucl-th/9308016 [nucl-th]].
	
	\bibitem{Henri:2020ezr}
	M.~Henri, O.~Lopez, D.~Durand, B.~Borderie, R.~Bougault, A.~Chbihi, Q.~Fable, J.~D.~Frankland, E.~Galichet and D.~Gruyer, \textit{et al.}
	Phys. Rev. C \textbf{101}, no.6, 064622 (2020)
	doi:10.1103/PhysRevC.101.064622.
	
	\bibitem{DiToro:1998ki}
	M.~Di Toro, V.~M.~Kolomietz and A.~B.~Larionov,
	Phys. Rev. C \textbf{59}, 3099-3108 (1999)
	doi:10.1103/PhysRevC.59.3099
	[arXiv:nucl-th/9807070 [nucl-th]].
	
	\bibitem{Wang:2020xgk}
	R.~Wang, Z.~Zhang, L.~W.~Chen, C.~M.~Ko and Y.~G.~Ma,
	Phys. Lett. B \textbf{807}, 135532 (2020)
	doi:10.1016/j.physletb.2020.135532
	[arXiv:2007.12011 [nucl-th]].

	\bibitem{PhysRevD.105.112002}
        Jiang, L. \textit{et al.} [Jefferson Lab Hall A 
        Phys. Rev. D \textbf{105}, 112002 (2022)
        doi:10.1103/PhysRevD.105.112002
        [arXiv:2203.01748 [nucl-ex]].


        \bibitem{PhysRevD.103.034604}
        L. Gu \textit{et al.} [Jefferson Lab Hall A],
        Phys. Rev. C \textbf{103}, 034604 (2021)
        doi:10.1103/PhysRevC.103.034604
        [arXiv:2012.11466 [nucl-ex]].

        \bibitem{Nature.560.617621}
        M. Duer \textit{et al.} [CLAS],
        Nature \textbf{560}, 617–621 (2018) 
        doi:10.1038/s41586-018-0400-z

        \bibitem{PhysLettB.608.4752}
        C. Barbieri, D. Rohe, I. Sick, and Louk Lapikas,
        Phys. Lett. B \textbf{608}, 47-52 (2005) 
        doi:10.1016/j.physletb.2004.12.072
        [arXiv:nucl-th/0411066].


 
	\bibitem{Lu:2019nmf}
	X.~Lu and J.~T.~Sobczyk,
	Phys. Rev. C \textbf{99}, no.5, 055504 (2019)
	doi:10.1103/PhysRevC.99.055504
	[arXiv:1901.06411 [hep-ph]].
	
	\bibitem{uboone_detector}
	R. Acciarri \textit{et al.} [MicroBooNE], 
	JINST 12, P02017 (2017)
	doi:10.1088/1748-0221/12/02/P02017.
	
	\bibitem{Phys.Rev.D.79.072002}
	A.A. Aguilar-Arevalo \textit{et al.} [MiniBooNE],
	Phys. Rev. D \textbf{79}, 072002 (2009)
	doi:10.1103/PhysRevD.79.072002
	[arXiv:0806.1449 [hep-ex]].
	
	\bibitem{Buss:2011mx}
	O.~Buss, T.~Gaitanos, K.~Gallmeister, H.~van Hees, M.~Kaskulov, O.~Lalakulich, A.~B.~Larionov, T.~Leitner, J.~Weil and U.~Mosel,
	Phys. Rept. \textbf{512}, 1-124 (2012)
	doi:10.1016/j.physrep.2011.12.001
	[arXiv:1106.1344 [hep-ph]].
	
	\bibitem{Mosel:2023zek}
	U.~Mosel and K.~Gallmeister,
	Phys. Rev. D \textbf{109}, no.3, 033008 (2024)
	doi:10.1103/PhysRevD.109.033008
	[arXiv:2308.16161 [nucl-th]].
	
	\bibitem{Song:2015hua}
	T.~Song and C.~M.~Ko,
	Phys. Rev. C \textbf{91}, no.1, 014901 (2015)
	doi:10.1103/PhysRevC.91.014901.
	
	\bibitem{Oset:1987re}
	E.~Oset and L.~L.~Salcedo,
	Nucl. Phys. A \textbf{468}, 631-652 (1987)
	doi:10.1016/0375-9474(87)90185-0.

        \bibitem{Gonzalez-Jimenez:2017fea}
        R.~Gonz\'alez-Jim\'enez, K.~Niewczas and N.~Jachowicz,
        Phys. Rev. D \textbf{97}, no.1, 013004 (2018)
        doi:10.1103/PhysRevD.97.013004
        [arXiv:1710.08374 [nucl-th]].

	\bibitem{MicroBooNE:2024xod}
	P.~Abratenko \textit{et al.} [MicroBooNE],
        Phys. Rev. D \textbf{110}, 013006 (2024)
        doi:10.1103/PhysRevD.110.013006
	[[arXiv:2402.19216 [hep-ex]].      

	\bibitem{Phys.Rev.Lett.131.101802.2023}	
	P. Abratenko \textit{et al.} [MicroBooNE], 
	Phys. Rev. Lett. 131, 101802 (2023)
        doi:10.1103/PhysRevLett.131.101802
	[[arXiv:2301.03706 [hep-ex]]].	

	\bibitem{Phys.Rev.Lett.128.151801.2022}
	P. Abratenko \textit{et al.} [MicroBooNE],
	Phys. Rev. Lett. 128, 151801 (2022)
        doi:10.1103/PhysRevLett.128.151801
        [arXiv2110.14023 [hep-ex]].

	\bibitem{arXiv2403.19574}
	P. Abratenko \textit{et al.} [MicroBooNE],
	[arXiv2403.19574 [hep-ex]].

	\bibitem{WSVD}
	W. Tang, X. Li, X. Qian, H. Wei, and C. Zhang, 
	JINST 12, P10002 (2017) 
        doi:10.1088/1748-0221/12/10/P10002
	[arXiv:1705.03568 [physics.data-an]].
 
 	\bibitem{arXiv2404.10948}
	P. Abratenko \textit{et al.} [MicroBooNE],
	[arXiv2404.10948 [hep-ex]].

    \bibitem{Phys.Rev.D.81.111102}
    Y. Kurimoto \textit{et al.} [SciBooNE],
    Phys. Rev. D \textbf{81}, 111102(R) (2010)
    doi:10.1103/PhysRevD.81.111102
    [arXiv:1005.0059 [hep-ex]].


    \bibitem{Phys.Lett.B.664.41}
    A.A. Aguilar-Arevalo \textit{et al.} [MiniBooNE],
    Phys. Lett. B 664, 41 (2008)
    doi:10.1016/j.physletb.2008.05.006
    [arXiv:0803.3423 [hep-ex]].
 
    \bibitem{Lalakulich:2012cj}
    O.~Lalakulich and U.~Mosel,
    Phys. Rev. C \textbf{87}, no.1, 014602 (2013)
    doi:10.1103/PhysRevC.87.014602
    [arXiv:1210.4717 [nucl-th]].

	\bibitem{Ericson:1988gk}
	T.~E.~O.~Ericson and W.~Weise,
	Clarendon Press, 1988,
	ISBN 978-0-19-852008-5.

	
    \bibitem{neut}
    Hayato, Y., Pickering, L. 
    Eur. Phys. J. Spec. Top. \textbf{230}, 4469–4481 (2021)
    doi:10.1140/epjs/s11734-021-00287-7
    [arXiv:2106.15809 [hep-ph]].

    \bibitem{genie}
    C. Andreopoulos, \textit{et al.},
    Nucl. Instrum. Meth. A \textbf{614}:87-104,2010
    doi:10.1016/j.nima.2009.12.009
    [arXiv:0905.2517 [hep-ph]].

     \bibitem{blockwise}
    S. Gardiner, 
    [arXiv:2401.04065 [hep-ex]].

    \bibitem{arXiv:2404.09949}
	P. Abratenko \textit{et al.} [MicroBooNE],
	[arXiv:2404.09949 [hep-ex]].
 
    \bibitem{arXiv:2406.10583}
	P. Abratenko \textit{et al.} [MicroBooNE],
    [arXiv:2406.10583 [hep-ex]]. 

      
    \bibitem{HERMES:2005mar}
	Airapetian, A. \textit{et al.} [HERMES],    
      Phys. Rev. Lett. \textbf{96}, 162301 (2006)
      [arXiv:hep-ex/0510030 [hep-ex]].
         
\end{thebibliography}
	\end{document}